\documentclass{article}

\usepackage{arxiv}

\usepackage[utf8]{inputenc} % allow utf-8 input
\usepackage[T1]{fontenc}    % use 8-bit T1 fonts
\usepackage{hyperref}       % hyperlinks
\usepackage{url}            % simple URL typesetting
\usepackage{booktabs}       % professional-quality tables
\usepackage{amsfonts}       % blackboard math symbols
\usepackage{nicefrac}       % compact symbols for 1/2, etc.
\usepackage{microtype}      % microtypography
\usepackage{lipsum}		% Can be removed after putting your text content
\usepackage{graphicx}
\usepackage{natbib}
\usepackage{doi}
\usepackage{bm}
\usepackage[varvw]{newtxmath}       % selects Times Roman as basic font
\usepackage{amsmath}
\mathchardef\mhyphen="2D % Define a "math hyphen"
\usepackage{booktabs}
\usepackage{multirow}
\usepackage{comment}

\title{Surrogate Modeling-Driven Physics-Informed Multi-fidelity Kriging: Path Forward to Digital Twin Enabling Simulation for Accident Tolerant Fuel}

\author{ 	{Kazuma ~obayashi} \\
	Nuclear Engineering and Radiation Science\\
	Missouri University of Science and Technology\\
	Rolla, MO 65409, USA \\
	\And
         {James ~Daniell } \\
	Nuclear Engineering and Radiation Science\\
	Missouri University of Science and Technology \\
\And
         {Shoaib ~Usman } \\
	Nuclear Engineering and Radiation Science\\
	Missouri University of Science and Technology \\
 \And
         {Dinesh ~Kumar } \\
	Department of Mechanical Engineering\\
	University of Bristol\\
	Bristol BS8 1TR, UK \\
 \And
      {Syed ~Alam} \\
	Nuclear Engineering and Radiation Science\\
	Missouri University of Science and Technology\\
	Rolla, MO 65409, USA \\
 }

\renewcommand{\shorttitle}{\textit{Handbook of Smart Energy Systems} }

\hypersetup{
pdftitle={A template for the arxiv style},
pdfsubject={q-bio.NC, q-bio.QM},
pdfauthor={David S.~Hippocampus, Elias D.~Striatum},
pdfkeywords={First keyword, Second keyword, More},
}

\begin{document}
\maketitle

\begin{abstract}
The Gaussian Process (GP)-based surrogate model has the inherent capability of capturing the anomaly arising from limited data, lack of data, missing data, and data inconsistencies (noisy/erroneous data) present in the modeling and simulation component of the digital twin framework \cite{kobayashi2022uncertainty,kobayashi2022practical,rahman2022leveraging,khan2022digital}, specifically for the accident tolerant fuel (ATF) concepts. However, GP will not be very accurate when we have limited high-fidelity (experimental) data. In addition, it is challenging to apply higher dimensional functions (>20-dimensional function) to approximate predictions with the GP. Furthermore, noisy data or data containing erroneous observations and outliers are major challenges for advanced ATF concepts. Also, the governing differential equation is empirical for longer-term ATF candidates, and data availability is an issue. Physics-informed multi-fidelity Kriging (MFK) can be useful for identifying and predicting the required material properties. MFK is particularly useful with low-fidelity physics (approximating physics) and limited high-fidelity data - which is the case for ATF candidates since there is limited data availability. This chapter explores the method and presents its application to experimental thermal conductivity measurement data for ATF. The MFK method showed its significance for a small number of data that could not be modeled by the conventional Kriging method. Mathematical models constructed with this method can be easily connected to later-stage analysis such as uncertainty quantification and sensitivity analysis and are expected to be applied to fundamental research and a wide range of product development fields. The overarching objective of this chapter is to show the capability of MFK surrogates that can be embedded in a digital twin system for ATF.
\end{abstract}

% keywords can be removed

\section{Introduction}

Whether the product is for commercial or industrial use, it is necessary to optimize its intended function under constraints (e.g., cost, durability, and efficiency) in developing the product. One factor contributing significantly to these limitations is the operating environment, especially in extreme environments such as nuclear fuels, where high temperatures and corrosive and radiation damage present more severe restrictions. Other harsh environments include space applications, particle accelerators, and fusion reactors. Undoubtedly, it is essential to research and develop innovative materials that can withstand these harsh environments. Previously, the corresponding author has published a number of works on harsh marine environment \cite{alam2019assembly,alam2019small1,alam2019small2,alam2019parametric,alam2019coupled,alam2020neutronic,alam2019neutronic,alam2020lattice,almutairi2022weight}, which can leverage ATF fuel.

%%%
In developing new materials such as ATF \cite{alam2019assembly,almutairi2022weight}, a procedure has been used in which each ATF property data is obtained through experiments, and then mathematical modeling is performed. One challenge with this process is that a certain number of sample data is required in order to ensure smaller statistical errors. This problem could be solved if given unlimited research funding and time. However, they are critical factors in the real world: preparation of experimental samples according to the material properties to be measured, setup of equipment, and measurement time. In order to solve this concern, modeling techniques using \textbf{physics-informed machine learning (ML)} have attracted attention.

In order to conceptualize digital twin (DT) technology, surrogate modeling is generally the first step since it can make the overall computational cost cheaper while making the predictions acceptably accurate - a win-win situation. Most of the previous DT surrogate model development have been utilizing \textbf{Kriging}, also known as \textbf{Gaussian process regression}. It is one of the ML-based modeling methods. Kriging model is based on the posterior distribution, and the distribution can be explained as a combination of the prior and the likelihood. A set of input vectors, weight parameters, and an observed target value are prerequisites for building a model with the Kriging. Advantages of this method include (1) the ability to calculate predictions and their variances and (2) the flexibility of model selection through kernel functions. As a precaution, kernel functions should be selected while carefully evaluating the model's predictive performance for each one. Also, the inverse of the variance-covariance matrix can be unstable if there are samples with the same input values but different output values in the training data. Therefore, the Gaussian process is competent when the quality and training data are sufficient.

As mentioned earlier, the Kriging is not optimal when insufficient data is available, as in the case of advanced materials research. \textbf{Multi-Fidelity Kriging (MFK)} is a method developed to overcome this issue. MFK is known as a data fusion technique that is used to combine two kinds of data: \textbf{low-fidelity} and \textbf{high-fidelity} data \cite{Chakraborty2021}. The classification of data is the followings:

\begin{itemize}
    \item {Low-fidelity (LF): approximation of the system behavior obtained by solving governing equations (i.e., differential equations and formulas) \cite{Chakraborty2021},}
    \item {High-fidelity (HF): accurate data obtained with real experiments.}
\end{itemize}

\noindent
The LF data is available (it can generate data more efficiently) in regards to the data acquisition since it is a physics-driven formula; however, that compromises the accuracy due to its inherent physics assumptions. In contrast, although highly accurate, the availability of obtaining HF data is challenging. The overall algorithm is to leverage the data fusion technique to combine both the LF and HD data for a trustworthy prediction of ATF properties. The MFK method can build a model by adding information obtained from the governing equations (LF) to a small number of HF data.

This chapter presents a mathematical explanation of the MFK and an example of applications to material property modeling. The material to be modeled was chosen to be uranium silicide ($\rm U_{3}Si_{2}$), a nuclear fuel material, and the experimental data reported by Shimizu \cite{Shimizu1965} was employed. In addition, the results of Kriging and MF K applied to the experimental data were compared, demonstrating the usefulness of MFK in innovative materials research.

\section{Multi-Fidelity Kriging (MFK)}

MFK is a multi-fidelity modeling method, and this chapter adopted the methodology from ``Surrogate Modeling Toolbox"\cite{SMT2019,bouhlel2016improved}. Now, let consider that we have $s$ levels of a real value function $z_{1}(x), \dots , z_{s}(x)$ and corresponding Kriging model $Z_{1}(x), \dots, Z_{s}(x)$, $x \in Q$ where $Q$ is a nonempty open set called the input parameter space. For the subscript, $i=1,\dots,s$, we define the order from the least accurate to the most accurate. In this section, we treat the case $s=2$ to simplify matters. 

\begin{equation}
    \begin{cases}
        Z_{2}(x) = \rho_{1}(x)\tilde{Z_{1}}(x) + \delta_{2}(x) \\
        \tilde{Z_{1}}(x) \perp \delta_{2}(x)
    \end{cases}
\end{equation} 

\noindent
where $\rho$ is a scaling/correlation factor, $\delta$ is a discrepancy function, and $\tilde{Z_{1}}(x)$ is a Gaussian process with distribution in the first level of a layer. Note: the experimental design sets, $\mathbf{D}_{1}$ and $\mathbf{D}_{2}$, must fulfill the relationship $\mathbf{D}_{2} \subseteq \mathbf{D}_{1}$. It can be interpreted as a method of numerical modeling based on the correlation between a Kriging model constructed by low-fidelity and high-fidelity data sets. 

Since this chapter focuses on applications of the MFK, please refer to the following references for detailed mathematical explanations: Kennedy and O'Hagan \cite{kennedy2001bayesian} and Le Gratiet \cite{le2013multi}. In addition, the basic Kriging method is explained by Rasmussen \cite{rasmussen2003gaussian} and its application to the nuclear field is shown by Kobayashi et al \cite{kobayashi2022practical}.

In addition to the MFK, we also introduce advanced models. There are methods that combine the MFK and partial least square  techniques. Partial Least Squares (PLS) is a statistical technique for analyzing the variations of an important variable in relation to underlying variables. The PLS method provides instructions (principal constituents) that maximize the variation of the quantity of interest \cite{SMT2019}. \textbf{MFK-PLS} and \textbf{MFK-PLSK} are methods that employ the PLS. These methods perform the PLS analysis on the high-fidelity to preserve the robustness to poor correlations between fidelity levels \cite{SMT2019, bouhlel2016improved}. These advanced models are expected to be more powerful when input variables are of higher dimensions.

\section{Material Property Modeling}

\begin{figure}[htbp]
    \centering
    \includegraphics[width=12cm]{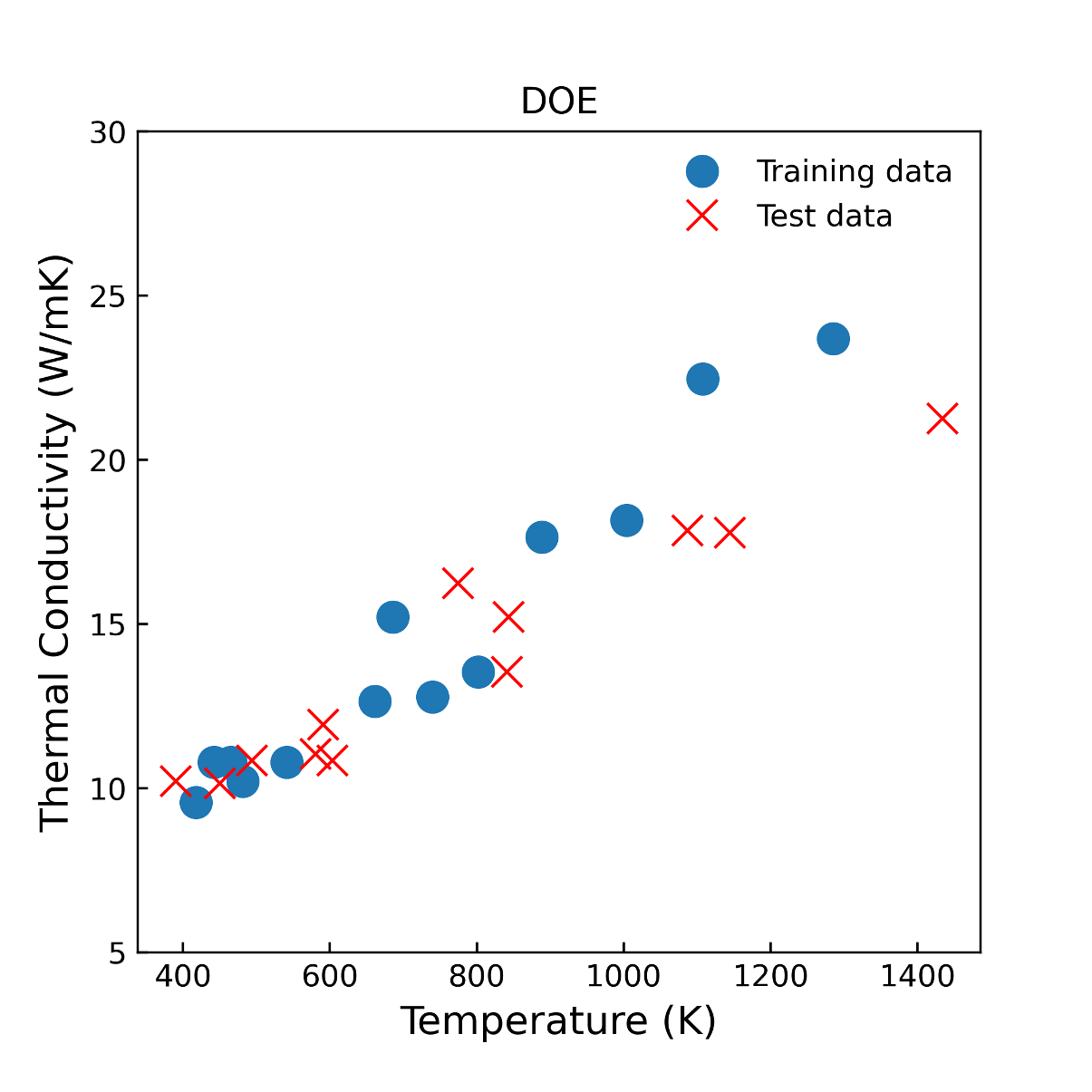}
    \caption{Experimental data reported by Shimizu \cite{Shimizu1965} and classification of subsets. The blue markers represent a training data set. The red cross markers a test data set.}
    \label{fig:dataset}
\end{figure}

Uranium silicide ($\rm U_{3}Si_{2}$) was selected for demonstrating the MFK. This material is reported to have a relatively higher density and thermal conductivity than a conventional uranium dioxide ($\rm UO_{2}$) fuel \cite{Gong2020b, Antonio2018, Metzger2017}. Hence, it has seemed to be a potential future nuclear fuel for light water reactors (LWRs) for several decades. One notable characteristic of the material is its thermal conductivity proportional to temperature rise. In this section, the Kriging and MFK are applied to actual thermal conductivity measurement data from the reference \cite{Shimizu1965}, respectively, and the differences in results are compared. The experimental data are randomly divided into two subsets: (1) training data and (2) test data. The training data is used to build a model, and the test one is prepared for evaluating a model (Fig. \ref{fig:dataset}).

\begin{figure}[htbp]
    \centering
    \includegraphics[width=16cm]{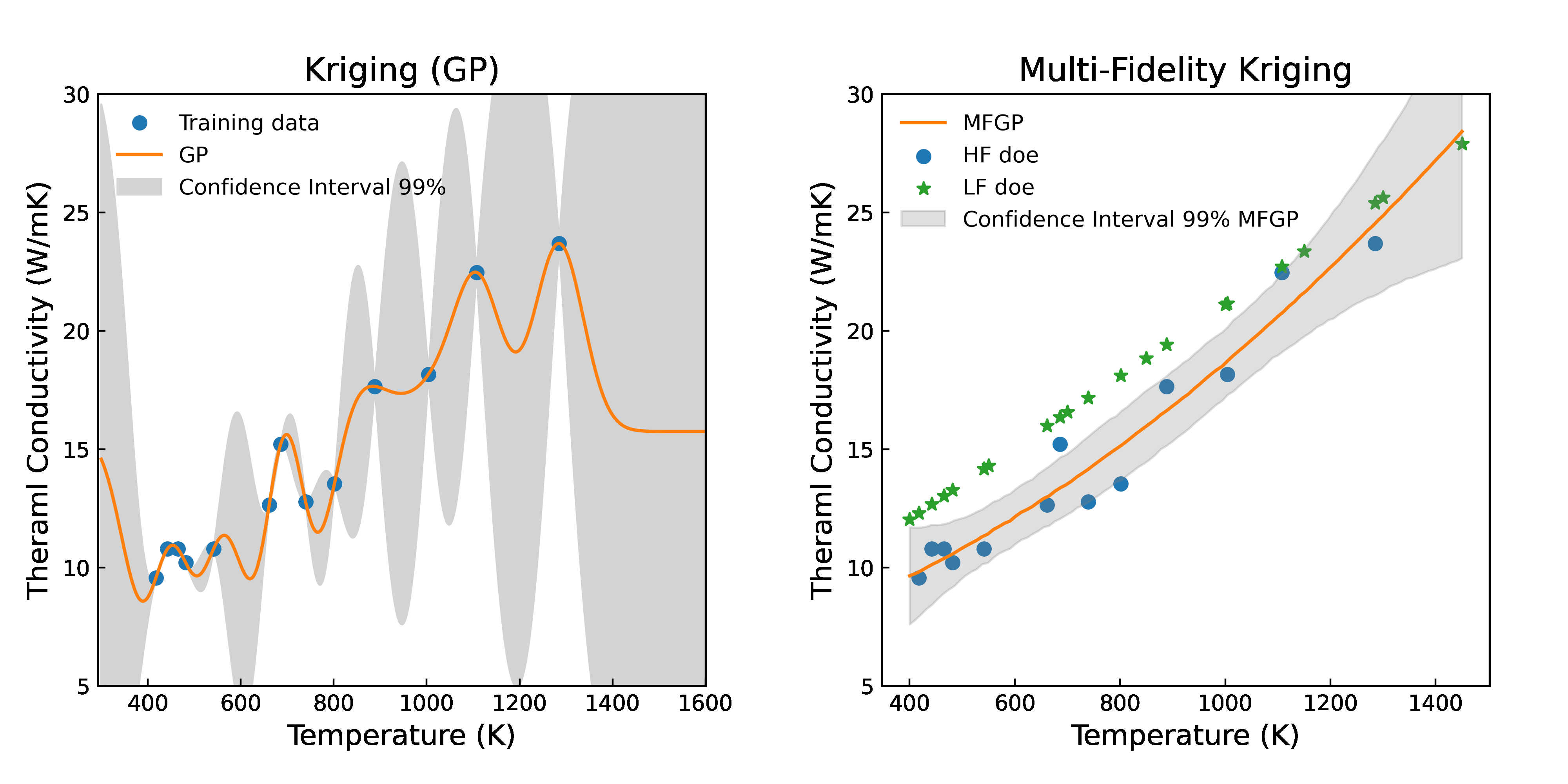}
    \caption{Results of the Kriging and the MFK. The left panel (Kriging) shows the expected values (orange line) have fluctuations and the confidence intervals between the training data points are very large. The right panel (MFK) shows the expected values are smooth following the LF (green) and its confidence intervals are converged.}
    \label{fig:gp_mfk}
\end{figure}

The left panel in Fig. \ref{fig:gp_mfk} shows the thermal conductivity model build with the Kriging. It reveals that the expected values are vibrating and confidence intervals are large. In particular, it can be concluded that the predictions in the regions below 400 K and above 1300 K, where no training data exist, are not meaningful from an empirical point of view. In addition, it can be read that the expected values have very large uncertainties in those regions. Therefore, the Kriging method is unsuitable for this demonstration's numerical modeling.

In order to apply the MFK, the LF must be defined. The thermal conductivity equation proposed by White et al. \cite{White2015} was adopted in this study. The thermal conductivity ($\lambda$) is described as a function of temperatures (T):

\begin{equation}
\label{eq:white}
    \lambda = 0.0151 \times \rm{T} + 6.004\,\,(\rm Wm^{-1}K^{-1}).
\end{equation}

\noindent
The LF dataset generated with Eq. \ref{eq:white} is represented as green markers in the right panel of Fig. \ref{fig:gp_mfk}. The model with the MFK was computed using the LF and the HF, and the result is shown by the orange line in the left panel of Fig. \ref{fig:gp_mfk}. It shows that the expected values are smooth and its confidence intervals are converged. A significant result is that, unlike the Gaussian process assumption method, where the predictions oscillate, MFK conveys the character of a positive slope in the LF data and yields more realistic predictions.

To quantitatively assess the validity of the two models, the value of the root-mean-square deviation (RMSD) method was computed using the test data. The RMSD can be expressed by the following:

%%%
\begin{equation}
\label{eq:rmsd}
    \text{RMSD} = \sqrt{\frac{\sum_{i=1}^{N}(\hat{y}_{i}-y_{i})^2}{N}}
\end{equation}
\noindent
where $N$ is the number of samples in the test dataset, $\hat{y}$ is the computed value with the MFK models, and $y$ is the input variable of the test dataset. The RMSD computed for Kringing, MFK, MFK-KPLS, and MFK-PLSK models are about 0.200, 0.176, 0.176, 0.176, respectively (see Fig. \ref{eq:rmsd}). It is revealed that the results of MFK-based models are almost equal in terms of the RMSD. However, the two models employing the PLS show more convergence in confidence intervals than the pure MFK model. Therefore, the models employing the PLS are superior in terms of reliability even for problems with one-dimensional input variables, as in the present example. We will expand this work to explore the feasibility of advanced signal processing algorithm techniques \cite{kabir2010non,kabir2010theory,kabir2010watermarking,kabir2010loss}  for advanced materials. Also,  data adjustments using Bayesian inference \cite{kumar2020nuclear}, data assimilation \cite{kumar2019influence}, uncertainty quantification \cite{kobayashi2022uncertainty}, and robust optimization \cite{kumar2020uncertainty, kumar2022multi,verma2022reliability,kobayashi2022data} will be performed to complete the studies. In order to obtain a better surrogate performance of MFK,  following the approach by Meng and  Karniadakis \cite{meng2020composite}, the team is developing a physics-driven and constitutive-based multi-fidelity approach, Physics-Informed Multi Fidelity Neural Network (PMNN) surrogate model  \label{fig:MFNN} to capture the complex thermomechanical behavior of ATF while identifying erroneous data within ATF systems. Our preliminary results show that  PMNN can capture new physics regimes combining low-fidelity and high-fidelity data/equations. We have built a volume swelling model of CVD SiC (1–6 dpa) based on \cite{kim2012material}, where high-fidelity experimental data were collected from \cite{katoh2014continuous} while considering low-fidelity approximate equation for volumetric swelling.

\begin{figure}[htbp]
    \centering
    \includegraphics[width=16cm]{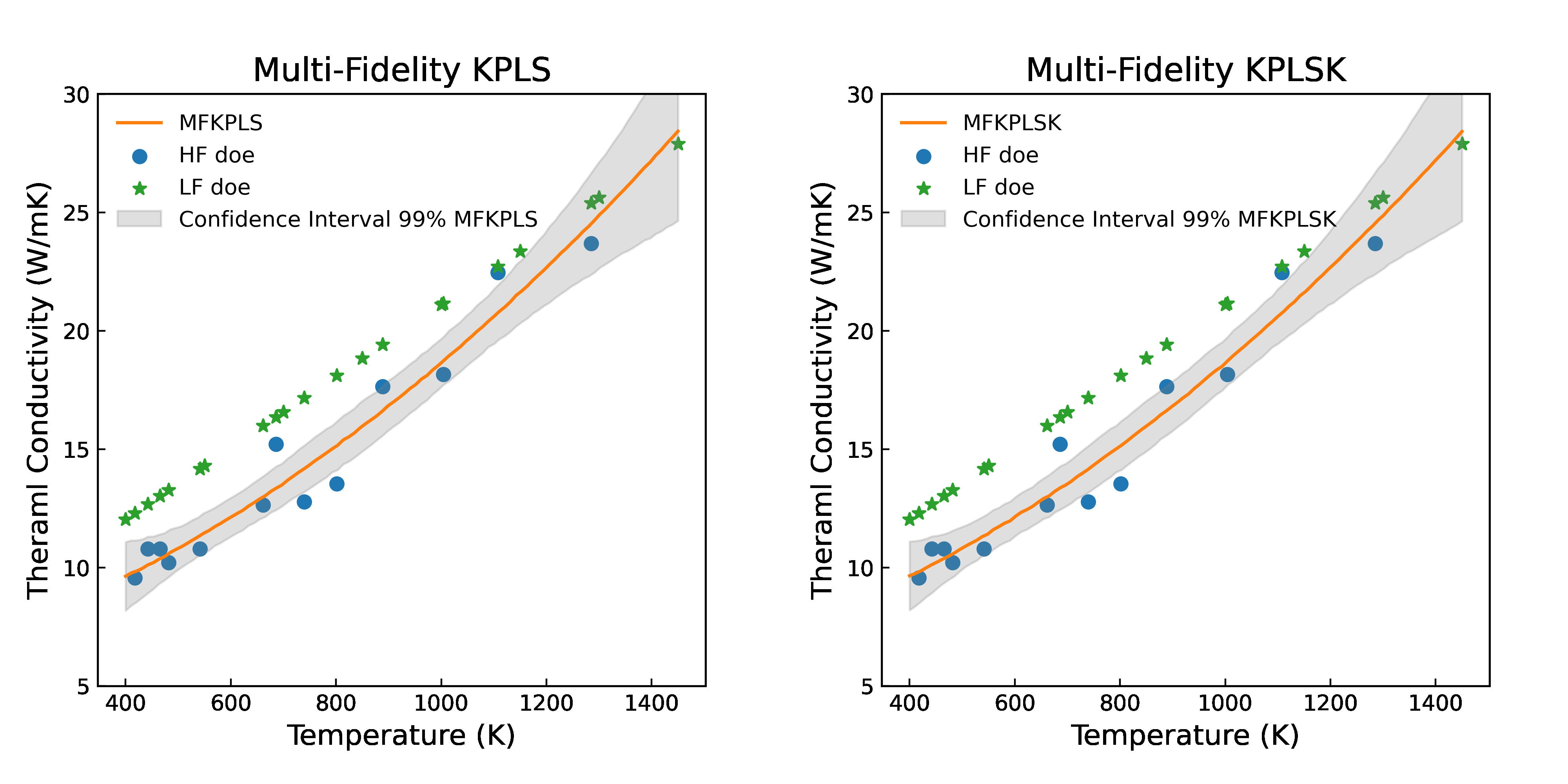}
    \caption{Results of the MFKPLS and MFKPLSK. The green markers represent the LF, and the blue dots are the HF data. The model computed with each method is represented with orange lines.}
    \label{fig:kplsk}
\end{figure}

\begin{figure}[htbp]
    \centering
    \includegraphics[width=12cm]{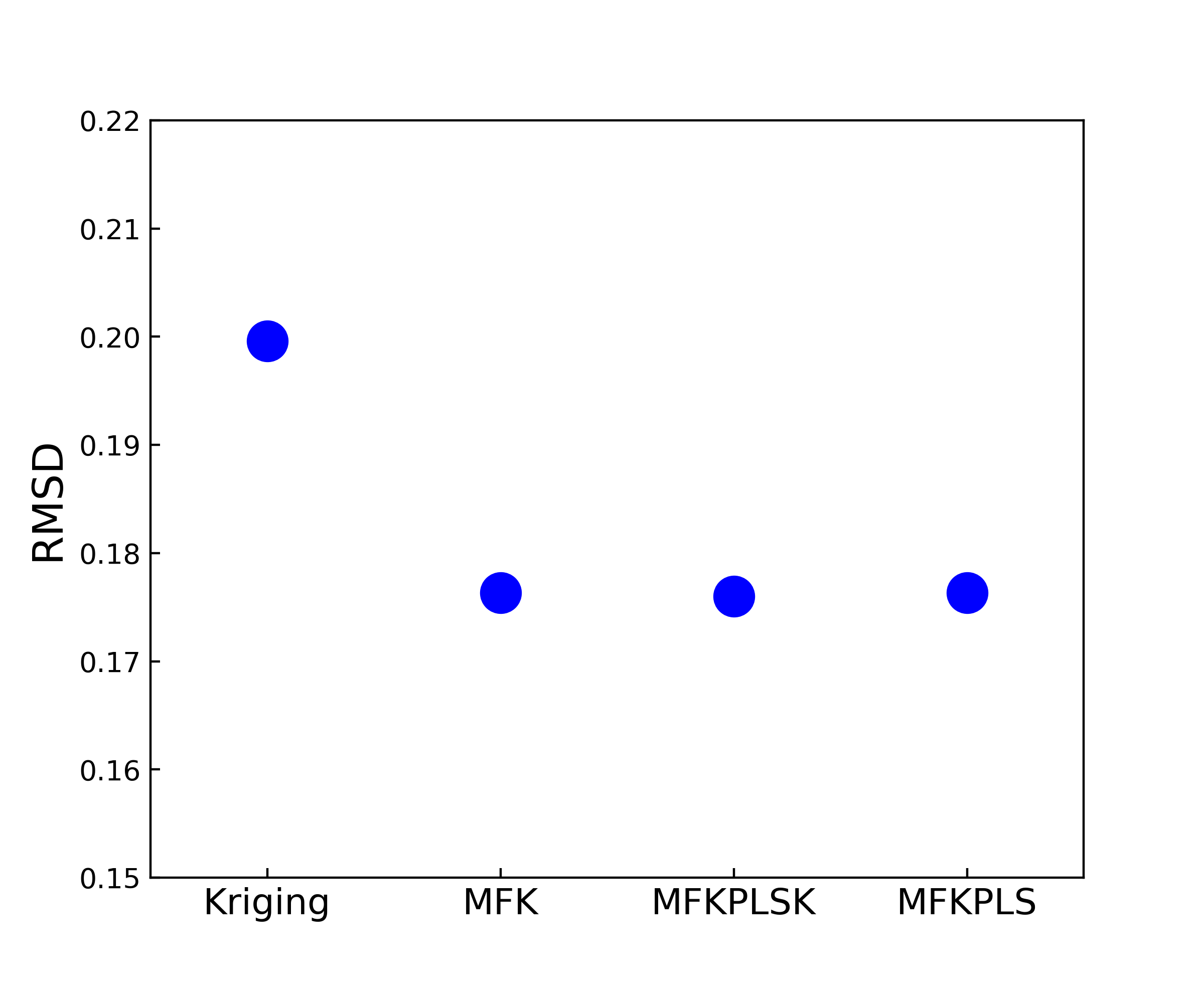}
    \caption{Computed RMSD for models. The value for Kriging is 0.200, and other models are about 0.176.}
    \label{fig:rmsd}
\end{figure}

\begin{figure}[htbp]
    \centering
    \includegraphics[width=14cm]{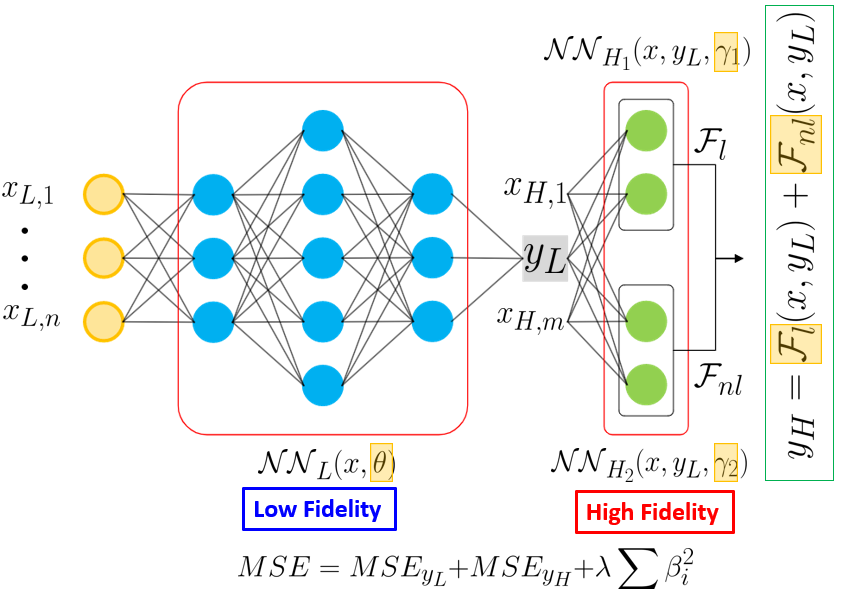}
    \caption{Physics-Informed Multi Fidelity Neural Network (PMNN) surrogate model (under development by following the approach \cite{meng2020composite}).}
    \label{fig:MFNN}
\end{figure}

% \cite{katoh2014continuous,kim2012material}
\begin{figure}[htbp]
    \centering
    \includegraphics[width=17cm]{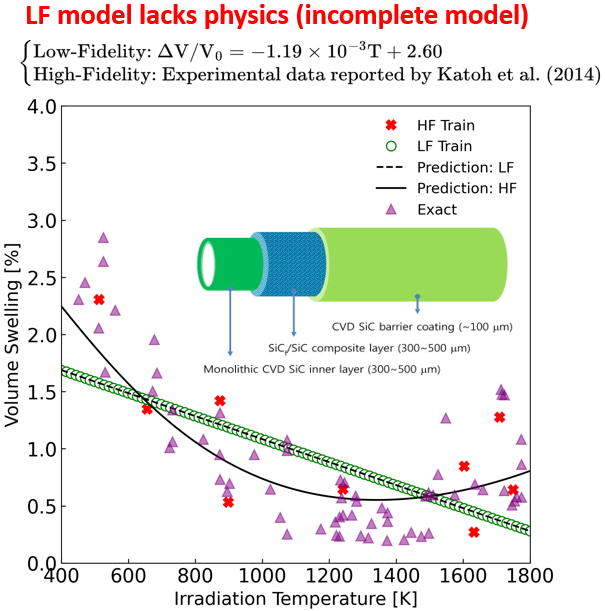}
    \caption{Volumetric swelling (\%) prediction leveraging Physics-Informed Multi Fidelity Neural Network (PMNN) surrogate model \cite{meng2020composite}) for volume swelling model of CVD SiC (1–6 dpa) based on \cite{kim2012material}, where high-fidelity experimental data were collected from \cite{katoh2014continuous} while considering low-fidelity approximate equation for volumetric swelling.}
    \label{fig:MFNN}
\end{figure}

\section{Conclusions}

Research and development of innovative materials are crucial in improving Digital Twin's performance and reliability. However, constraints such as the cost and time required to test the material have created a need for more efficient modeling methods of experimental data. In this chapter, Multi-Fidelity Kriging (MFK) was introduced as one of the solutions as a Digital Twin surrogate and modeling the thermal conductivity of nuclear fuel was presented as an example of as a use case. As a result, the MFK succeeded in building more reliable models for problems that cannot be modeled by the Kriging. The strength of the model constructed with this method is that predictions and their reliability (uncertainty) can be calculated. This statistical information is indispensable for product design uncertainty quantification, sensitivity analysis, and multi-scale analysis. The MFK is a promising analytical method for the development of innovative materials and is expected to find further applications in industry, space engineering, and energy. In the future, we will explore the feasibility of digital twins for advanced materials.

\section*{Acknowledgement}
The computational part of this work was supported in part by the National Science Foundation (NSF) under Grant No. OAC-1919789.

\bibliographystyle{unsrtnat}
\bibliography{references}  %%% Uncomment this line and comment out the ``thebibliography'' section below to use the external .bib file (using bibtex) .





% \end{thebibliography}

\end{document}